\documentclass[12pt]{article}

\newcommand{\mult}[3]{({\bf{#1}},\,{\bf{#2}},\,{\bf{#3}})}

\newcommand{\p}[1]{(\ref{#1})}

\newcommand{\nn}{\nonumber}
\newcommand{\be}{\begin{equation}}
\newcommand{\ee}{\end{equation}}
\newcommand{\bea}{\begin{eqnarray}}
\newcommand{\eea}{\end{eqnarray}}
\newcommand{\ba}{\begin{array}}
\newcommand{\ea}{\end{array}}
\newcommand{\disty}{\displaystyle}
\newcommand{\bet}{\bar\eta}

\newcommand{\bth}{{\bar\theta}}
\newcommand{\bD}{\bar D}

\newcommand{\brho}{{\bar\phi}}
\newcommand{\bpsi}{\bar\psi}
\newcommand{\bxi}{\bar\xi}
\newcommand{\Hv}{H_{,v\,}}
\newcommand{\Hr}{H_{,\phi\,}}
\newcommand{\Hbr}{H_{,\brho\,}}
\newcommand{\Hvv}{H_{,vv\,}}
\newcommand{\Hvr}{H_{,v\phi\,}}
\newcommand{\Hvbr}{H_{,v\brho\,}}

\newcommand{\Gv}{G_{,v\,}}
\newcommand{\Gr}{G_{,\phi\,}}
\newcommand{\Gbr}{G_{,\brho\,}}
\newcommand{\Gvv}{G_{,vv\,}}

\newcommand{\Grbr}{G_{,\phi\brho\,}}

\newcommand{\vf}{\varphi}
\newcommand{\bvf}{{\bar\vf}}

\newcommand{\Fvv}{\F_{,\,vv}\hspace{0.1pt}}

\newcommand{\Fvvv}{\F_{,\,vvv}}
\newcommand{\Fvvf}{\F_{,\,vv\vf}}
\newcommand{\Fvvbf}{\F_{,\,vv\bvf}}

\newcommand{\Fvvvv}{\F_{,\,vvvv}}
\newcommand{\Fvvvf}{\F_{,\,vvv\vf}}
\newcommand{\Fvvvbf}{\F_{,\,vvv\bvf}}
\newcommand{\Fvvff}{\F_{,\,vv\vf\vf}}
\newcommand{\Fvvbfbf}{\F_{,\,vv\bvf\bvf}}

\newcommand{\Fvf}{\F_{,\,v\vf}}
\newcommand{\Fvbf}{\F_{,\,v\bvf}}

\newcommand{\F}{{\tt F}}
\title{Hyper-K\"ahler geometries and nonlinear supermultiplets}
\author{\v{C}.~Burd\'ik$^{1}$, S.~Krivonos$^{2}$, A.~Shcherbakov$^{2}$}
\date{}
\begin{document}
\maketitle
\begin{center}
$^{1}$ Department of Mathematics, Czech Technical University, Trojanova 13,\\
120 00 Prague 2, Czech Republic\\
$^{2}$ Bogoliubov Laboratory of Theoretical Physics, JINR,\\
141980 Dubna, Russia
\end{center}
\begin{abstract}
\noindent
It is presented a method of construction of sigma-models with target space geometries different from conformally flat ones.
The method is based on a treating of a constancy of a coupling constant as a dynamical constraint following as an equation of motion.
In this way we build~$N=4$ and~$N=8$ supersymmetric four-dimensional sigma-models in~$d=1$ with hyper-K\"ahler target space
possessing one isometry, which commutes with supersymmetry.
\end{abstract}
\section*{Introduction}
One-dimensional theories (i.e. mechanics) with four and eight supercharges stand out among all one-dimensional
theories with extended supersymmetry. This is due to existence of linear~$N=4$ and~$N=8$ irreducible representations
having no auxiliary fields~\cite{Gates}. Construction of sigma-model actions corresponding to these models turns out
to be quite easy task to do. A detailed analysis of sigma-model geometries of arising bosonic manifolds was performed
in papers~\cite{GPS,N48} and revealed an interesting fact: under quite general assumptions concerning the
structure of sigma-model actions arising bosonic manifolds are to be conformally flat.
This is a direct evidence that those considerations seem to overlook some points relating to other possible geometries,
because a dimensional reduction of four-dimensional~$N=2$ sigma-models~\cite{Alvarez} down to~$d=1$ is known to lead
to hyper-K\"ahler bosonic manifolds. It is quite easy to understand what was missed: in one dimension there exists a
wide class of \emph{nonlinear} off-shell supermultiplets. These are the very supermultiplets that play a crucial role
in constructing supersymmetric sigma-models with different type of target space geometry~\cite{{BuKS},{KS},{BKS2},{DI}}.
Unlike to linear supermultiplets, construction and classification of nonlinear off-shell ones are much more difficult
problem which is complicated by their absence in higher dimensions.

In this paper we describe a dualization of a coupling constant procedure to construct a nonlinear realization of supersymmetry
that allows us to build the most general $N=4,8$ supersymmetric four-dimensional sigma-models with one triholomorphic isometry.

The idea of the coupling constant dualization can be most easily demonstrated by an example of a conformal mechanics
which is governed by the action
\be\label{1}
 S= \int {\tt d} t \left[ \dot x^2 - \frac{g^2}{x^2} \right]
\ee
with a bosonic field~$x(t)$ depending on time~$t$ only and~$g$ being a coupling constant. The constraint~$g=const$ can obviously
be interpreted as a solution to a differential equation
\be\label{2}
\frac{{\tt d} g}{{\tt d} t} =0.
\ee
Thus in such an approach we have got a system~\p{1} with a constraint~\p{2}. Alternative to solving the constraint~\p{2} is
including it into the action~\p{1} with a Lagrange multiplier~$\phi(t)$
\be\label{3}
S= \int {\tt d} t \left[ \dot x^2 -\frac{g^2}{x^2}  - 2 \phi \dot{g} \right]
\ee
with a quantity~$g(t)$ being no more a constant but some function of~$t$. Varying \p{3} over $\phi$ we will get just \p{2},
while  the ``equation of motion'' for $g$ reads
\be\label{4}
g= x^2 \dot\phi\;.
\ee
Eliminating the ``coupling constant''~$g$ we get the following action
\be\label{5}
S= \int {\tt d} t \left[ \dot x^2 + x^2 \dot\phi^2\right],
\ee
which is easily recognized as the action of a $D=2$ free particle written in the polar coordinates.

\section{Constructing~$N=4$ hyper-K\"ahler $\sigma$-manifold}
In this section we construct an~$N=4$ supersymmetric $\sigma$-model with a hyper-K\"ahler geometry of its scalar manifold
using dualization of a coupling constant in a way described above. Since the coupling constant dualization increases the number
of the physical scalars by one, therefore we should start from a model with at least a three-dimensional target space to
have a four-dimensional one as a result. Appropriate three-dimensional model is based on a linear~\mult341 supermultiplet
with four scalars (three physical and one auxiliary) and four fermions~\cite{IvSm,BerPash,IKLech}.

We consider a supersymmetry algebra with four odd generators. Appropriate superspace~${\bf R}^{1|2}$ may be equipped with
covariant spinor derivatives which satisfy the following relations
$$\{ D, \bar D \} = 2i\frac{{\tt d}\phantom{t} }{{\tt d} t} , \qquad \{D,D \} =  \{\bar D,\bar D \} =0.$$
In terms of~$N=2$ superfields the supermultiplet~\mult341 is given by a real~$N=2$ superfield~$V(t,\theta,\bth)$, a
chiral~$N=2$ one~$\Phi(t,\theta,\bth)$ and its conjugated~$\bar\Phi(t,\theta,\bth)$
\be\label{drhochir}
\bar V = V, \qquad D \bar\Phi = \bD \Phi = 0.
\ee
To maintain~$N=4$ supersymmetry this~$N=2$ formulation has to be augmented by an extra~$N=2$ supersymmetry
transformation which mixes the  the superfields~$V$ and~$\Phi$
\be\label{dsusytr}
\delta V = - \epsilon \bD \bar\Phi - \bar\epsilon D \Phi, \qquad
\delta \Phi = \epsilon \bD V,\qquad
\delta \bar\Phi = \bar\epsilon D V.
\ee
The standard~$N=4$ supersymmetric $\sigma$-model action for the supermultiplet~\mult341 looks like~\cite{IvSm}
\be\label{S0}
S_1 = \int {\tt d}t {\tt d}^2\theta\;  G\, \left( D V \bD V + D \Phi \bD \bar\Phi \right),
\ee
with a metric~$G$ being an arbitrary function of the superfields~$V, \Phi,\bar\Phi$. The action~\p{S0} is invariant
with respect to the additional~$N=4$ supersymmetry transformations~\p{dsusytr}. To apply the above mentioned dualization
one should have a constant, but we still miss it. To overcome this problem we just add a potential term
\be\label{dS2}
S_2 = g \int {\tt d}t {\tt d}^2\theta  H(V,\Phi,\bar\Phi), \qquad g = \mbox{const}
\ee
to the $\sigma$-model action~$S_1$. The potential term contains a dimensional constant~$g$ and is invariant under
manifest~$N=2$ supersymmetry. To enlarge it to the~$N=4$ one, we should require the invariance of~$S_2$ under the
transformations~\p{dsusytr}. This results in restriction on the function~$H$ to be a harmonic one
\be\label{dLaplace}
H_{VV} + H_{\Phi\bar\Phi} = 0.
\ee
Therefore,~$N=4$~$d=1$ supersymmetric action we will deal with acquires the following form
\be\label{dS}
S = S_1 + S_2 = \int {\tt d}t {\tt d}^2\theta
    \left[\makebox[0pt]{$\phantom{\disty\frac12}$} G \left( D V \bD V + D \Phi \bD \bar\Phi \right)
    + g H  \right] \,
\ee
with the function~$H$ satisfying Laplace equation~(\ref{dLaplace}). After integration over the Grassmann coordinates
we get the component form of the action
\be\label{dScomp1}
\ba{l}
\disty S = \int {\tt d} t \left[ \makebox[0pt]{$\phantom{\disty\frac12}$}
    -\left( \Gvv + \Grbr \right) \psi\bpsi \xi\bxi
    + i\dot v \left( \Gr \xi\bpsi + \Gbr \bxi \psi \right)
    + g \left( \Hvr \xi\bpsi + \Hvbr\psi \bxi \right)-\right.\\
\disty \phantom{S=\int d}\left.
    - A \left( \Gv \left( \psi\bpsi - \xi\bxi \right) + \Gr \xi \bpsi - \Gbr \bxi\psi + g \Hv\right)
    + g \Hvv \left( \psi\bpsi - \xi \bxi \right)-\right.\\
\disty \phantom{S=\int d}\left. -i \dot\phi \left(\Gr \left( \psi\bpsi - \xi\bxi \right)
    - 2 \Gv \psi\bxi \right)
    + i \dot\brho\left(\Gbr \left( \psi\bpsi - \xi\bxi \right)
    - 2 \Gv \xi\bpsi \right)+\right.\\
\disty \phantom{S=\int d}\left.
    + G \left( \dot v^2 + 4 \dot\phi \dot\brho + A^2 + i \dot\psi \bar\psi - i \psi \dot{\bar\psi}
    + i \dot\xi \bar\xi - i \xi \dot{\bar\xi}\right)
    - i g \left( \Hbr \dot\brho - \Hr\dot\phi \right)
    \makebox[0pt]{$\phantom{\disty\frac12}$}\right].
\ea
\ee
Here the components of the superfields~$V$ and~$\Phi$ are defined as follows
$$
\ba{cll}
\disty v(t)=V, & \disty \quad \psi(t)=DV, \quad \bpsi(t)=-\bD V,
    & \disty \quad A(t) = \frac{1}{2}\left[ D, \bD \right] V,\\
\disty \phi(t)=\Phi,\quad \bar\phi(t) = \bar\Phi, & \disty \quad \xi(t)=D\Phi,\quad \bxi(t)=-\bD \bar\Phi &
\ea
$$
with the right hand sides being evaluated at~$\theta=\bar\theta=0$.
As it was previously described, to dualize the constant~$g$ we just add a term, which provides
the constancy of~$g$, to the action~(\ref{dScomp1})
\be\label{LagrMult}
S \quad \to \quad S - \int {\tt d}t\, y(t) \dot g .
\ee
In contrast to the previously considered example with a conformal mechanics, now the coupling constant~$g$ is
involved into the action~$S$ only linearly. Thus, its interpretation now is as a Lagrange multiplier for the some
constraint on the field content of our theory which includes now one additional bosonic field $y(t)$. It is easy to see
that this additional constraint expresses  the auxiliary field~$A$ in terms of the $y(t)$
\be\label{eq1}
A =\frac1{\Hv} \left[ \Hvv \left(\psi\bpsi - \xi\bxi \right) - i \left( \Hbr \dot\brho-\Hr \dot\phi\right) - \dot y
    +\Hvr \xi\bpsi + \Hvbr \psi \bxi\;\right].
\ee
Therefore, the number of the physical scalars is increased and  we arrived at a supermultiplet~\mult440.
It consists of the for physical bosons~$v,\phi,\bar\phi$ and~$y$ and four fermions~$\psi,\bar\psi$ and~$\xi, \bar\xi$.
With respect to the full~$N=4$ supersymmetry they transform as follows
\bea\label{tr1}
&& \delta v = \eta\bpsi - \bet \psi + \epsilon\bxi - \bar\epsilon\xi, \quad
\delta\phi = -\bet \xi - \epsilon\bpsi, \quad
\delta\brho = \eta \bxi + \bar\epsilon\psi, \nn\\
&& \delta y = -i\eta ( \Hv \bpsi + \Hbr \bxi)
    - i \bet ( \Hv \psi + \Hr \xi )
    + i \epsilon ( \Hv \bxi - \Hr \bpsi )
    + i\bar\epsilon ( \Hv \xi - \Hbr\psi), \nn\\
&& \delta\psi = -i\eta \dot v + \eta A + 2i \epsilon \dot\brho, \quad
\delta\bpsi = i\bet \dot v + \bet A - 2i \bar\epsilon \dot\phi, \\
&& \delta\xi = -i\epsilon \dot v - \epsilon A - 2i \eta \dot\phi, \quad
\delta\bxi = i\bar\epsilon \dot v - \bar\epsilon A + 2i \bet \dot\brho, \nn
\eea
where~$\eta$ and~$\epsilon$ are the supersymmetry parameters and expression for~$A$ is given by formula~\p{eq1}.

Now we see the main distinction of this new \mult440 supermultiplet from the known ones: transformations~\p{tr1} are
highly nonlinear and involve an arbitrary harmonic function~$H$. As well as the known ones, the constructed nonlinear
supermultiplet is defined \emph{off-shell}.

Substituting the expression for the auxiliary field~$A$ back into the action~\p{LagrMult} we get
\be\label{finAct}
S=\int {\tt d} t \left[ G \left( \dot v^2 + 4 \dot\phi \dot\brho \right) +
    \frac{G}{\Hv^2} \Bigl(\dot y - i \Hr \dot\phi + i \Hbr \dot\brho \Bigr)^2
    +\mbox{fermions} \right].
\ee
Kinetic part of this action describes a metric of a $\sigma$-model manifold
$${\tt d} s^2 = G \left( {\tt d} v^2 + 4 {\tt d}\phi\, {\tt d}\brho \right) +
    \frac{G}{\Hv^2} \Bigl({\tt d} y - i \Hr {\tt d}\phi + i \Hbr {\tt d}\brho \Bigr)^2.$$
The Weyl tensor constructed for this metric is different from zero, so that this manifold is genuinely not conformally-flat.
Moreover, imposing an additional requirement
\be\label{dGH}
G = \Hv
\ee
we get a Ricci--flat bosonic manifold with a general Gibbons--Hawking metric for a hyper-K\"ahler manifold with
one triholomorphic isometry~\cite{GH}. Under the condition~\p{dGH} the action gets the form
$$
\ba{l}
\disty S=\int {\tt d}t \left[ \Hv \left( \dot v^2 + 4 \dot\phi \dot\brho \right) +
    \frac1{\Hv} \Bigl(\dot y - i \Hr \dot\phi + i \Hbr \dot\brho \Bigr)^2 \right.\\
\disty \phantom{S=\int {\tt d}}  +\left.
    \psi\bar\xi \left( i\Hvbr \dot v - 2i \Hvv \dot\phi
    - i\frac{ \Hr\Hvr \dot\phi - \Hbr\Hvbr \dot\brho}{\Hv} + \dot y \frac{\Hvbr}{\Hv}\right)\right.\\
\disty \phantom{S=\int {\tt d}}  +\left.
    \xi\bar\psi \left( -i\Hvr \dot v + 2i \Hvv \dot\brho
    - i\frac{ \Hr\Hvr \dot\phi - \Hbr\Hvbr \dot\brho}{\Hv} + \dot y \frac{\Hvr}{\Hv}\right)\right.\\
\disty \phantom{S=\int {\tt d}}  + \left.
    \left( \psi\bar\psi - \xi\bar\xi \right)\left( i\dot\phi \left( \Hvr - \frac{\Hvv\Hr}{\Hv} \right)
    -i\dot\brho \left( \Hvbr - \frac{\Hvv\Hbr}{\Hv} \right) + \dot y \frac{\Hvv}{\Hv}\right)\right]
\ea
$$
with four-fermionic terms disappearing, as it should be for Ricci-flat $\sigma$-model manifolds.

Thus, presented method of dualization allowed us to construct a nonlinear supermultiplet with component structure~\mult440
and find the action based on this supermultiplet. The latter corresponds to a not conformally flat manifold and becomes
a hyper-K\"ahler one if condition~\p{dGH} holds.

Let us add that the expression for the auxiliary field~$A$~\p{eq1} coincides (up to a total time-derivative term) with that one previously found in~\cite{KS},
while the idea of construction of an auxiliary field through a general superspace potential term  was firstly proposed by
E.~Ivanov~\cite{private}.

\section{Constructing~$N=8$ hyper-K\"ahler $\sigma$-manifold}

This section is based on an~$N=4$ superfield formalism therefore we first introduce notations. A superspace~${\bf R}^{1|4}$ we are dealing with is parameterized by one even coordinate~$t$ and four odd coordinates~$\theta_i$ and~$\bar\theta^i$
with the index~$i$ being~$SU(2)$ one running~$i=1,2$. All superfields to be considered are supposed to live in this superspace.
To single out an irreducible representation from a general superfield we make use of covariant spinor derivatives~$D^i$
and~$\bar D_i$ defined on~${\bf R}^{1|4}$ and satisfying the following relations
$$\{D^i, {\bar D}_j \} = 2\,i\, \delta^i_j\, \frac{{\tt d} }{{\tt d} t},\qquad \{D^i, D^j \}=\{\bar D_i, \bar D_j \} =0.$$
In a full analogy with $N=4$ supersymmetric case we will start from an irreducible $N=8$ supermultiplet with three physical bosons,
eight fermions and five auxiliary bosons, i.e.~({\bf 3},{\bf 8},{\bf 5}) multiplet. Such a representation is
described~\cite{ABC} in~${\bf R}^{1|4}$ by a real~$N=4$ superfield~$V(t,\theta,\bar\theta)$ and a chiral~$N=4$
superfield~$\Phi(t,\theta,\bar\theta)$
\be\label{irrep}
D^i \Phi = \bar D_i \bar\Phi = 0, \quad D^i D_i V = \bar D^i \bar D_i V =0.
\ee
The constraints \p{irrep} leave among the components  of the superfields~$V$ and~$\Phi$ the following  independent ones:
$$%\be
\ba{ccl}
\disty v(t)=V, & \disty \psi_i(t) = - i \bar D_i V, \quad \bar\psi^i(t) = - i D^i V, &
\disty A_{ij}(t) = i[\bar D_{(i}, D_{j)}]V,\\[0.5ex]
\disty \varphi(t)=\Phi, & \disty \xi_i(t) = - i \bar D_i \Phi, \quad \bar\xi^i(t) = - i D^i \bar\Phi,&
\disty B(t)= D^i D_i\bar\Phi.
\ea
$$%\ee
The right hand sides of the above expressions are supposed to be taken with vanishing~$\theta_i$ and~$\bar\theta^i$.

One should note that the constraints~(\ref{irrep}) impose the following restrictions on the superfield $V$~\cite{IKL}
\be
\frac{\partial}{\partial t} [D^i , \bar D_i ] V=0 \quad \Rightarrow \quad [D^i , \bar D_i ] V = 2 g, \qquad  g = {\tt const}.
\ee
If $g \neq 0$ it appears in the $\theta$'s decomposition of the superfield~$V$
\be
V(t,\theta,\bar\theta) = v(t) + i \theta_i \bar\psi^i(t) + i \bar\theta^i \psi_i(t)
    + \frac12 \theta^i \bar\theta^j \left( i A_{ij}(t) - \varepsilon_{ij} g \right) + \ldots
\ee
Having these~$N=4$ superfields one can easily build a supersymmetric action as an integral of a real superfunction over the whole superspace
\be\label{action}
S=\int {\tt d} t L = -\int {\tt d} t {\tt d}^2 \theta {\tt d}^2 \bar\theta{}\, \F(V,\Phi,\bar\Phi).
\ee
Being constructed in terms of manifest~$N=4$ superfields the action~(\ref{action}) is just  ~$N=4$ supersymmetric, not~$N=8$.
To promote it to~$N=8$ one should require the action to be invariant with respect to an additional~$N=4$ supersymmetry
\be\label{hidden}
\delta V = \eta_i D^i \bar\Phi + \bar\eta^i \bar D_i \Phi, \quad
\delta \Phi = - \eta_i D^i V, \quad \delta \bar\Phi = - \bar\eta^i \bar D_i V.
\ee
This additional $N=4$ supersymmetry commutes with the manifest one and extends it to $N=8$.
The invariance of the action \p{action} with respect \p{hidden} puts the restricting the prepotential~$\F$ to be a harmonic function
\be\label{Laplace}
\frac{\partial^2 \F}{\partial V \partial V} + \frac{\partial^2 \F}{\partial \Phi \partial \bar\Phi} = 0.
\ee
Finally, after  performing Grassmann integration in eq.~(\ref{action}) one gets the following expression for the
Lagrangian\footnote{The bilinear products~$\psi^2$ and~$\bar\psi^2$ stands for~$\psi^i\psi_i$ and~$\bar\psi_i\bar\psi^i$ respectively.}
\be\label{Lagr_compon}
\ba{l}
L=\Fvv \left( \dot v^2 + 4\dot\varphi \dot{\bar\varphi}
    -i \dot\psi^i \bar\psi_i + i \psi^i \dot{\bar\psi_i}
    -i \dot\xi^i \bar\xi_i + i \xi^i \dot{\bar\xi_i} \right)\\[0.3ex]
\phantom{L=}
-\frac14 \Fvvv \left(\psi^2 \bar\psi^2 + \xi^2 \bar\xi^2 - 4 \xi^i \psi_i \bar\xi_j \bar\psi^j \right)
    -\frac12 \Fvvff \xi^2 \bar\psi^2 - \frac12 \bar\xi^2 \psi^2\\[0.3ex]
\phantom{L=}
    -\frac12 \Fvvvf \left( \bar\psi^2 \xi^i \psi_i - \xi^2 \bar\xi_i \bar\psi^i \right)
    -\frac12 \Fvvvbf \left( \psi^2 \bar\xi_i \bar\psi^i - \bar\xi^2 \xi^i \psi_i \right)\\[0.3ex]
\phantom{L=}
-i \dot v \left( \Fvvf \xi^i \bar\psi_i - \Fvvbf \psi^i \bar\xi_i \right)
    +i \left( \Fvvf \dot\varphi- \Fvvbf \dot{\bar\varphi} \right) \left( \psi^i \bar\psi_i
    - \xi^i \bar\xi_i\right)\\[0.3ex]
\phantom{L=}
    + 2 i \Fvvv \left( \dot\varphi \bar\xi_i \psi^i -\dot{\bar\varphi} \xi_i \bar\psi^i\right)\\[0.3ex]
\phantom{L=}
    + \Fvv \left( \frac18 A^{ij} A_{ij} + \frac14 B \bar B - \frac14 g^2 \right)
-i g \left( \Fvf \dot\varphi - \Fvbf \dot{\bar\varphi} \right)\\[0.3ex]
\phantom{L=}
-\frac12 \left( i A_{ij} + \varepsilon_{ij} g \right)
    \left( \Fvvv \left( \psi^j \bar\psi^i - \xi^j \bar\xi^i \right) + \Fvvf \xi^j \bar\psi^i
    + \Fvvbf \psi^j \bar\xi^i \right)\\[0.3ex]
\phantom{L=}
-\frac14 B \left( \Fvvbf \psi^2 - 2 \Fvvv \xi^i \psi_i - \Fvvf \xi^2 \right)\\[0.3ex]
\phantom{L=}
-\frac14 \bar B \left( \Fvvf \bar\psi^2 - 2 \Fvvv \bar\xi_i \bar\psi^i - \Fvvbf \bar\xi^2 \right)
\ea
\ee
The Lagrangian~\p{Lagr_compon} still contains the auxiliary fields which have to be eliminated by their
equations of motion. But before doing this first let us turn to the constant~$g$ which, as one can see
from~(\ref{Lagr_compon}), serves as a coupling constant corresponding to an interaction of a particle with a
background electromagnetic field.

Following the prescription given above, we incorporate constancy of the coupling constant~$g$ into the action
\be\label{action_mod}
S \to S - \int {\tt d} t \;  y(t)\dot g(t)
\ee
with the help of a Lagrange multiplier~$y(t)$.
Now we are ready to eliminate the all set of auxiliary fields~$A_{ij}(t)$,~$B(t)$ and~$g(t)$ from the modified
action~(\ref{action_mod}) using their equations of motion. This results in  the action written
in terms of physical fields only
\be\label{action_final}
S=\int {\tt d} t\; \bigl[ K - U \bigr]
\ee
with the kinetic term equal to
$$
\disty K=\Fvv \left( \dot v^2 + 4\dot\varphi \dot{\bar\varphi}
    -i \dot\psi^i \bar\psi_i + i \psi^i \dot{\bar\psi_i}
    -i \dot\xi^i \bar\xi_i + i \xi^i \dot{\bar\xi_i} \right)
    + \frac{\bigl( \dot y - i \Fvf\dot\varphi + i\Fvbf\dot{\bar\varphi}\bigr)}{\Fvv}^2
$$
and potential one
$$
\ba{l}
\disty U = \frac14 \left(\psi^2 \bar\psi^2 + \xi^2 \bar\xi^2 - 4 \xi^i \psi_i \bar\xi_j \bar\psi^j\right)
    \left( \Fvvvv + \frac{\Fvvf\Fvvbf-2\Fvvv^2}{\Fvv}  \right)\\[0.3ex]
\disty \phantom{U=}
+ \frac12 \xi^2 \bar\psi^2 \left( \Fvvff - \frac{3 \Fvvf^2}{\Fvv}\right)
    + \frac12 \bar\xi^2 \psi^2 \left( \Fvvbfbf - \frac{3 \Fvvbf^2}{\Fvv}\right)\\[0.3ex]
\disty \phantom{U=}
+ \frac12 \left( \bar\psi^2 \xi^i \psi_i - \xi^2 \bar\xi_i \bar\psi^i \right)
    \left(\Fvvvf - \frac{3\Fvvv\Fvvf}{\Fvv}\right)\\[0.3ex]
\disty \phantom{U=}
+ \frac12 \left( \psi^2 \bar\xi_i \bar\psi^i - \bar\xi^2 \xi^i \psi_i \right)
    \left(\Fvvvbf - \frac{3\Fvvv\Fvvbf}{\Fvv}\right)
    - 2 i \Fvvv \left( \dot\varphi \bar\xi_i \psi^i -\dot{\bar\varphi} \xi_i \bar\psi^i\right)\\[0.3ex]
\disty \phantom{U=}
+i \dot v \left( \Fvvf \xi^i \bar\psi_i - \Fvvbf \psi^i \bar\xi_i \right)
    -i \left( \Fvvf \dot\varphi- \Fvvbf \dot{\bar\varphi} \right) \left( \psi^i \bar\psi_i
    - \xi^i \bar\xi_i\right)\\[0.3ex]
\disty \phantom{U=}
-\frac{ \dot y - i \Fvf\dot\varphi + i\Fvbf\dot{\bar\varphi}}{\Fvv}
    \left( \Fvvv  \left( \psi^i \bar\psi_i - \xi^i \bar\xi_i\right) + \Fvvf \xi^i \bar\psi_i
    + \Fvvbf \psi^i \bar\xi_i\right).
\ea
$$
The kinetic term defines the following metric of the bosonic manifold
\be\label{metric}
{\tt d}\, s^2 =\Fvv \left( {\tt d} v^2 + 4{\tt d}\varphi {\tt d}{\bar\varphi} \right)
    + \frac1{\Fvv} \bigl( {\tt d} y - i \Fvf{\tt d}\varphi + i\Fvbf{\tt d}{\bar\varphi}\bigr)^2.
\ee
The metric~(\ref{metric}) is of Gibbons--Hawking form~\cite{GH} corresponding to the most general
four-dimensional hyper-K\"ahler manifold with one triholomorphic isometry, which is realized as a shift along the coordinate~$y$.

Thus, in a such simple manner we construct $N=8$ supersymmetric hyper-K\"ahler $\sigma$-model.

\section*{Conclusion}
In the paper we presented a simple idea for constructing~$N=4$ and~$N=8$ supersymmetric hyper-K\"ahler $\sigma$-models by
dualizing a coupling constant, which may either be present in the superfield decomposition or serve as an coupling constant
for the potential term.

The idea of such a dualization is based on an ambivalent interpretation of coupling constants in one dimensions:
on the one hand it is just a constant, on the other hand~-- it may be interpreted as some constant values of angular momenta.
Dualized system contains one additional scalar field and describes a mechanics with an arbitrary value of such momenta.
In particular, dualization turns the tensor
supermultiplet into the nonlinear hypermultiplet for the case of~$N=4$.
The most essential point is that transformation properties of constructed hypermultiplet is nonlinear.
Moreover, in the case of~$N=4$ dualization includes one additional harmonic function and gives a nonlinear hypermultiplet defined off-shell
whose transformations properties under supersymmetry crucially depend on this function.
The case of~$N=8$ is a little bit different and the question whether the hypermultiplet is off-shell requires more detailed study.

There are some questions yet to be solved: it unclear dualization of what constants is essential, when constructed nonlinear
supermultiplets are defined off-shell or on-shell, how to describe such supermultiplets in terms of superfield approach, etc.

\vspace{1cm}

The authors are grateful to E.~Ivanov, D.~Sorokin and M.~Vasiliev for valuable and useful discussions.
This work was partially supported by grants RFBR-06-02-16684, DFG~436 Rus 113/669/0-3 and GACR~201/05/0857.

\end{document}